\documentclass{article}
\usepackage{amsmath}
\usepackage{amssymb}
\usepackage{cite}

\def\vep{\varepsilon}

\def\bea{\begin{eqnarray}}
\def\eea{\end{eqnarray}}

\newcommand{\be}{\begin{equation}}
\newcommand{\ee}{\end{equation}}
\newcommand{\beqn}{\begin{eqnarray}}
\newcommand{\eeqn}{\end{eqnarray}}
\newcommand{\beqnn}{\begin{eqnarray*}}
\newcommand{\eeqnn}{\end{eqnarray*}}

\begin{document}
\newcommand{\pst}{\hspace*{1.5em}}

\begin{center} {\Large \bf
{Justification of the ``symmetric damping'' model 
\\
of the dynamical Casimir effect in a cavity 
\\
with a semiconductor mirror
}\footnote{Dedicated to Margarita Alexandrovna Man'ko on the occasion of her
70th birthday.}
}
 \end{center}



\begin{center} {\bf
Victor V. Dodonov
}\end{center}


\begin{center}
{\it
Instituto de F\'{\i}sica, Universidade de Bras\'{\i}lia,\\
 Caixa Postal 04455, 70910-900 Bras\'{\i}lia, DF, Brazil
}

$^*$e-mail:~~~~vdodonov@fis.unb.br\\
\end{center}

\begin{abstract}\noindent
A ``microscopic'' justification of the ``symmetric damping'' model of a quantum 
oscillator with time-dependent frequency and time-dependent damping is given. This model is used to predict results of experiments
on simulating the dynamical Casimir effect in a cavity with a photo-excited semiconductor mirror. It is shown that
the most general bilinear time-dependent coupling of a selected oscillator (field mode) to a bath of harmonic oscillators results in
two {\em equal\/} friction coefficients for the both quadratures, provided all the coupling coefficients are proportional to
a single arbitrary function of time whose duration is much shorter than the periods of all oscillators.
The choice of coupling in the {\em rotating wave approximation\/} form leads to the ``mimimum noise'' model
of the quantum damped oscillator, introduced earlier in a pure phenomenological way.

\end{abstract}



\section{Introduction}
\pst

Semiconductor mirrors have been used in the laser physics since the
first years of the laser era \cite{ElMan67,Carm64,Sooy65,Birn65}.
Recently, they have found once more application. Namely, it is quite
probable that they will help to realize a long-standing dream of an
experimental verification of the so called Dynamical Casimir Effect (DCE).
The story of this effect began 40 years ago, when Moore \cite{Moore} 
showed that the motion of ideal mirrors
could result in a creation of quanta of the electromagnetic field from
the initial vacuum state. 
Since then, this exciting phenomenon
was a subject of numerous theoretical studies
(see reviews in \cite{D-rev1,DD-rev2,DCas60}). 

The effect seemed to be extremely small, since the velocities of
boundaries achievable in a laboratory are much less than the velocity of light.
For this reason, it was thought for a long time that hardly the DCE could be observed.
The situation began to change in 1990s, when
a possibility of an enhancement of the effect inside cavities with {\em oscillating\/} boundaries under the conditions of
parametric resonance was discovered \cite{DKM90,DK96,Lamb,Dal99,Plun00}.
But exciting high frequency oscillations of rigid walls (about few GHz or higher) and maintaining  them for a sufficiently long time
(about $1\,$s) \cite{DK96} is still a great experimental challenge.

However, the task can be simplified, if one recognizes that all schemes that could be
used to create quanta from the initial vacuum state of some circuit or cavity are based, after all, 
on a parametrical change of the resonance frequency of the system. 
But this can be achieved not only by changing the geometry (moving walls),
but by changing other parameters, 
for example, the inductance and capacitance of some quantum circuit or properties of some Josephson junction
\cite{DOM89,Man91}.
Nowadays this idea has been rediscovered and embodied in different practical schemes 
\cite{Ciuti07,Taka08,Taka09,Sasha,Johan09,Ciuti09}.

At the same time the idea of simulating ``nonadiabatic Casimir effect'' and other quantum
phenomena using a medium with a rapidly
 decreasing in time refractive index (``plasma window'') was formulated by Yablonovitch \cite{Yabl89}.
Different theoretical schemes based on fast changes of the carrier concentration 
in semiconductors illuminated by laser pulses were discussed in \cite{Yabl89-2,Oku95,Loz},
and a possibility of creating an effective semiconductor {\em microwave mirror\/} in this way
was confirmed experimentally \cite{Padua}. The further development was
a proposal (named ``MIR experiment'')  \cite{Padua05} to simulate a motion
of a boundary, using an effective electron-hole ``plasma mirror,''
created periodically on the surface of a semiconductor slab
(attached to some part of the superconducting wall of a high-$Q$ cavity)
by illuminating it with a sequence of short laser pulses.
The amplitude of an effective displacement of the boundary is determined in such a case by the
thickness of the semiconductor slab, which can be made up to few
millimeters, resulting in a relative change of the cavity eigenfrequency up to $10^{-2}$,
instead of the maximal possible value $10^{-8}$  for mechanically driven mirrors
(this limit is due to tremendous internal stresses arising inside a material if
the frequency of the surface vibrations belongs to the GHz band \cite{DK96}).
The current status of the MIR experiment was reported in \cite{Pad-Cas60}.
A possibility to enhance and control the {\em static\/} Casimir force by illuminating semiconductor surfaces
was shown in \cite{Inui04,Chen07}.

Note that the thickness of the photo-excited conducting layer on the surface of the semiconductor slab is much smaller
than the thickness of the slab itself. It is determined by the absorption coefficient of laser radiation,
so it is about few micrometers or less, depending on the laser wavelength. Therefore laser pulses with the surface
energy density about few $\mu$J/cm$^{2}$ can create a highly conducting layer with the carrier concentration
exceeding $10^{17}\,$cm$^{-3}$, which gives rise to an almost maximal possible change of the cavity eigenfrequency 
for the given geometry. 
On the other hand, the conductivity of the layer is not extremely high due to a moderate value of the mobility
in the available materials (such as highly doped GaAs), which is about $1\,$m$^2$V$^{-1}$s$^{-1} $ \cite{Pad-Cas60}.
For this reason, effects of dissipation during the excitation-recombination process inside the layer
 cannot be neglected, since they can change the picture drastically \cite{DD-rev2}. 

A model taking into account the dissipation was introduced in \cite{DD-rev2,D-job05}.
It was based on the assumption that the effects of dissipation can be described within the frameworks
of the linear Heisenberg--Langevin operator equations of the form ($\hbar=1$)
\be
d\hat{x}/dt = \hat{p} -\gamma_x(t)\hat{x} +\hat{F}_x(t),
\qquad
d\hat{p}/dt = -\gamma_p(t)\hat{p} -\omega^2(t)\hat{x} +\hat{F}_p(t),
\label{Fx-Fp}
\ee
where $\hat{x}$ and $\hat{p}$ are dimensionless quadrature operators of the selected
mode of the EM field. These operators are
normalized in such a way that the mean number of photons equals 
${\cal N}=\frac12\langle \hat{p}^2 + \hat{x}^2 -1\rangle$.
Two noise operators 
$\hat{F}_x(t)$ and $\hat{F}_p(t)$ with zero mean values (commuting with $\hat{x}$ and $\hat{p}$)
are necessary to preserve the canonical commutator 
$\left[\hat{x}(t),\hat{p}(t)\right] = i$. 
These operators were assumed to be delta-correlated,
\be
\langle \hat{F}_j(t) \hat{F}_k(t')\rangle =\delta(t-t')\chi_{jk}(t),
\qquad j,k = x,p.
\label{sigjk}
\ee

The system of linear equations (\ref{Fx-Fp})  can be solved
explicitly for arbitrary time-dependent functions $\gamma_{x,p}(t)$,
$\omega(t)$ and $\hat{F}_{x,p}(t)$. 
It appears that the commutation relation $[\hat{x}(t), \hat{p}(t)]=i$ can be preserved {\em exactly\/}
under the condition $\chi_{xp} - \chi_{px}= 2i\gamma(t)$ for any function
$\gamma(t)=\left(\gamma_p+\gamma_x\right)/2$  \cite{DD-rev2, D-job05}.
The ``symmetric damping'' model with $\gamma_x =\gamma_p$ was chosen in \cite{DD-rev2,D-job05}
mainly due its simplicity. Later, the generic case of $\gamma_x \neq \gamma_p$ was studied 
in \cite{DD-JRLR06,Blaub,D-PRA09}. It was shown that there are some phenomenological reasons to connect the
damping and noise coefficients as follows: $\chi_{xp} = - \chi_{px}= i\gamma(t)$,
$\chi_{xx}=\gamma_x G$, $\chi_{pp}=\gamma_p G$, where the factor $G\ge 1$ depends on the reservoir temperature.
However, the concrete value of the ``asymmetry coefficient''
$y=\left(\gamma_p -\gamma_x\right)/\left(\gamma_p +\gamma_x\right)$ remained as a free parameter.
Although the choice $y=0$ permits one to simplify immensely many formulas \cite{D-PRA09},
 it seems that there is no possibility to deduce the value of $y$ from some general
phenomenological principles \cite{Blaub}, except for ``aesthetic'' ones.

On the other hand, the physical results, such as the rate of photon generation,
 can depend significantly on the value of $y$ \cite{DD-JRLR06}. This is because all formulas
for the physical quantities contain a special solution of the classical nonstationary oscillator
equation $\ddot\vep + \omega_{ef}^2(t) \vep =0$ with 
$\omega_{ef}^2(t) =\omega^2(t) + \dot{\delta}(t) -\delta^2(t)$, where
$ \delta(t)=\left[\gamma_x(t) - \gamma_p(t)\right]/2$.
If the damping coefficient $\gamma(t)$ is much smaller than the amplitude of variations of the
cavity eigenfrequency $\omega(t)$ and if this coefficient varies slowly with time (or it is constant),
then the corrections to the effective frequency, $\dot{\delta}(t)$ and $\delta^2(t)$, can be
safely neglected. This case was considered in \cite{D-PRA98}, where it was shown explicitly that only
the total damping coefficient $\gamma$ is important if $\gamma_x=const$, $\gamma_p=const$ and
$\gamma \ll \omega$. But in the case of the MIR experiment the total damping coefficient varies
very fast, in the time scale of the order of the recombination time, which must be less than $30\,$ps
\cite{DD-rev2}, whereas the period of the field mode is about $400\,$ps. Moreover, the
maximum value of $\gamma$ in this case is only twice smaller than the amplitude of variations of
$\omega(t)$. Therefore the term $\dot{\delta}(t)$ cannot be neglected, and it gives rise to 
significant changes in comparison with the symmetric case of $y=\delta(t)=0$ \cite{Blaub,D-PRA09}.

In principle, the value of $y$ (as well as of all the damping and noise coefficients)
could be deduced from some rigorous model, which would take into account
explicitly (i) the coupling of the field mode with
electron--hole pairs inside the semiconductor slab, 
(ii) the coupling of electrons and holes with phonons or other quasiparticles, 
responsible for the damping mechanisms, and (iii) a time dependence of the
number of carriers, which disappear after a short recombination time.
Unfortunately, it seems that no model of this kind is available now.
Therefore I consider in this paper a surrogate ``microscopic'' model, where 
the real dissipative system is replaced by a set of harmonic oscillators ({\em bosonic\/} reservoir)
and the real interactions are replaced 
by an effective quadratic bilinear bosonic Hamiltonian.
A justification is that this simple model was used in many text books  when
the authors considered field modes in non-ideal cavities \cite{Louis73,ScuZub97,Weiss99,GZ00}. Actually,
it was considered in numerous publications for a long period since at least 1950s
\cite{Toda58,Toda59,Magal59,George60,GWL63,FKM65,GH65,Uller66,Moll68,Glaub69,Agar71,OCon88}. 
The difference from the numerous existing studies
is that here I consider {\em the most general\/} bilinear coupling {\em with time-dependent coefficients}.
In the overwhelming majority of other papers only some specific couplings were considered,
such as the so-called RWA (Rotating Wave Approximation)  coupling \cite{GWL63,GH65,Moll68,Glaub69,OCon88},
the coordinate--coordinate coupling \cite{Toda58,Toda59,Magal59,FKM65,Uller66,OCon88}, the momentum--coordinate
coupling \cite{Toda58,OCon88}, a sum of coordinate--coordinate and momentum--coordinate couplings \cite{Agar71},
 but all the coefficients
in the interaction Hamiltonians were supposed to be {\em time-independent\/}.

The scheme adopted in this paper was introduced in \cite{Kor} and developed in \cite{DMM95}. 
It was used for studying the decoherence and transfer of quantum states between modes in 
the Fabry--Perot cavity with resonantly oscillating boundaries in \cite{DAM05}.
The next section is devoted to a generalization to the case of arbitrary time-dependent
coupling coefficients. In Sec. \ref{sec-short} I show how the two equal damping coefficients
$\gamma_x=\gamma_p$ arise naturally in the special case of short-time interaction
(which corresponds perfectly to the conditions of the MIR experiment), if all coupling coefficients
are proportional to a {\em single\/} function of time (which can be connected with the total
number of carriers inside the photo-excited semiconductor film).
The ``minimum noise'' model arises naturally under the additional assumption of the RWA coupling.

\section{Oscillator coupled to nonstationary bosonic bath}
\label{sec-gen}

Let us assume that the ``central'' oscillator with the time-dependent frequency
$\omega(t)$, described in terms of the ``coordinate'' $x_0$ and canonically conjugated ``momentum'' $p_0$ 
(representing the selected field mode in the cavity) is coupled to a
large number of the ``bath'' oscillators with constant frequencies $\omega_i$,
and coordinates $x_i$ and momenta $p_i$.  The Hamiltonian of the whole (closed) system has the form
\be
\hat{H}= \frac 12\left[\hat{p}_0^2+\omega^2(t)\hat{x}_0^2\right] + \frac 12 \sum_{i=1}^N
\left(\hat{p}_i^2+\omega_i^2\hat{x}_i^2\right)
+\sum_{i=1}^N \left(z_i\hat{p}_i\hat{p}_0+v_i\hat{p}_i\hat{x}_0
+u_i\hat{x}_i\hat{p}_0+g_i\hat{x}_i\hat{x}_0\right),
\label{5.1}
\end{equation}
where coupling coefficients $z_i$, $v_i$, $u_i$, $g_i$ can be arbitrary functions of time.
The effective masses of every oscillator are assumed to  be equal to unity (this can be achieved by rescaling
the coordinates).
It is convenient to introduce the $(2N+2)$-dimensional vector ${\bf q}=({\bf Q},\xi )$,
whose components are defined as follows:
${\bf Q}= \left(p_0, x_0\right)$, 
$\xi =\left(p_1, p_2,\ldots, p_N,x_1, x_2,\ldots,x_N\right)$.

The simplest description of the evolution of quantum systems with
multidimensional quadratic Hamiltonians, such as (\ref{5.1}), can be achieved in the
\emph{Wigner representation} \cite{Wigner,book}.
The time-dependent Wigner function of the whole system can be written as
\begin{equation}
W(\mathbf{q},t)=\int G(\mathbf{q},\mathbf{q^{\prime },}t)W(\mathbf{q}
^{\prime },0)d\mathbf{q}^{\prime }  \label{w1}
\end{equation}
where the propagator $G(\mathbf{q},\mathbf{q}^{\prime },t)$ satisfies the
initial condition $G(\mathbf{q},\mathbf{q}^{\prime },0)=\delta (\mathbf{q}-
\mathbf{q}^{\prime })$. One of many remarkable properties of the Wigner function is
that its evolution in the case of quadratic Hamiltonians is governed by the
\emph{first-order\/} partial differential equation \cite{book,Moyal}
(whereas the
Schr\"{o}dinger equation or the equation for the \emph{density matrix\/}
contain the \emph{second} order derivatives).
Therefore, the whole propagator is reduced to a
delta-function \cite{Moyal,Taka54,KruPof78,Rud80,Gad89},
$G(\mathbf{q},\mathbf{q}^{\prime },t)=\delta (\mathbf{q}-\mathbf{q}_
{\ast }(t;\mathbf{q}^{\prime }))$,  
where vector
$\mathbf{q}_{\ast }(t;\mathbf{q}^{\prime })
=\mathbf{R}(t) \mathbf{q}^{\prime }$
is the solution to the classical equations of motion satisfying the
initial condition $\mathbf{q}_{\ast }(0;\mathbf{q}^{\prime })=\mathbf{q}
^{\prime }$.
Recall that for systems with quadratic Hamiltonians,
the Heisenberg equations of motion for
operator $\widehat{\mathbf{q}}$ have exactly the same form as the equations
for classical generalized coordinates (or the first-order mean values),
so $\mathbf{R}(t)$ is a {\em symplectic\/} $(2N+2)\times (2N+2)$ matrix
satisfying the linear equation
\begin{equation}
\dot {{\bf R}}= {\cal A}{\bf R}, \qquad {\bf R}(0)={\bf I}_{2N+2},
\label{3.5}
\end{equation}
where ${\bf I}_{M}$ is the $M\times M$ unity matrix.
It is useful to split the matrices $\mathbf{R}$ and ${\cal A}$
 into rectangular blocks in accordance with the structure of vector ${\bf q}=({\bf Q},\xi )$:
\begin{equation}
\mathbf{R}=\left\|
\begin{array}{cc}
\mathbf{R}_{11} & \mathbf{R}_{12} \\
\mathbf{R}_{21} & \mathbf{R}_{22}
\end{array}
\right\| ,
\qquad
{\cal A}=\left\|
\begin{array}{cc}
{\cal A}_{11} & {\cal A}_{12} \\
{\cal A}_{21} & {\cal A}_{22}
\end{array}
\right\| .    
\label{w5}
\end{equation}
Then
Eq.  (\ref{3.5}) is equivalent to the 
following set of equations:
\begin{equation}\dot {{\bf R}}_{11}={\cal A}_{11}{\bf R}_{11}+{\cal A}_{
12}{\bf R}_{21},\qquad {\bf R}_{11}(0)={\bf I},\label{5.2}\end{equation}
\begin{equation}\dot {{\bf R}}_{21}={\cal A}_{21}{\bf R}_{11}+{\cal A}_{
22}{\bf R}_{21},\qquad {\bf R}_{21}(0)=0,\label{5.3}\end{equation}
\begin{equation}\dot {{\bf R}}_{12}={\cal A}_{11}{\bf R}_{12}+{\cal A}_{
12}{\bf R}_{22},\qquad {\bf R}_{12}(0)=0,\label{5.4}\end{equation}
\begin{equation}\dot {{\bf R}}_{22}={\cal A}_{21}{\bf R}_{12}+{\cal A}_{
22}{\bf R}_{22},\qquad {\bf R}_{22}(0)={\bf I}.\label{5.5}\end{equation}
For Hamiltonian (\ref{5.1}), the matrices ${\cal A}_{jk}$ have the following
explicit forms:
\begin{equation}
{\cal A}_{11}=\left\Vert\begin{array}{cc}
0&-\omega^2(t)\\
1&0\end{array}
\right\Vert ,
\qquad
{\cal A}_{22}=
\left\Vert\begin{array}{cc}
0&-\mbox{diag}\left(\omega_1^2,\ldots ,\omega_i^2,\ldots\right)\\
\mbox{diag}(1,\ldots ,1,\ldots )&0\end{array}
\right\Vert ,
\label{5.6}
\end{equation}
\begin{equation}
{\cal A}_{12}=\left\Vert\begin{array}{cccccccc}
-v_1&\cdots&-v_i&\cdots&-g_1&\cdots&-g_i&\cdots\\
z_1&\cdots&z_i&\cdots&u_1&\cdots&u_i&\cdots\end{array}
\right\Vert ,
\qquad
{\cal A}_{21}=\left\Vert\begin{array}{cc}
-u_1&-g_1\\
\vdots&\vdots\\
-u_i&-g_i\\
\vdots&\vdots\\
z_1&v_1\\
\vdots&\vdots\\
z_i&v_i\\
\vdots&\vdots\end{array}
\right\Vert .
\label{5.8}
\end{equation}
The notation $\mbox{diag}\left(a_1,\ldots ,a_i,\ldots\right)$
means the diagonal matrix with the elements $a_1,\ldots ,a_i,\ldots$.

I assume that the initial total Wigner function is factorized
with respect to the $\mathbf{Q}$ and $\xi$ variables 
and that the initial Wigner function of the reservoir is \emph{Gaussian\/}
 (hereafter $\hbar=1$)
\begin{equation}
W(\mathbf{q},0 )=W_{0}(\mathbf{Q})W_{1}(\xi ), \qquad
W_{1}(\xi )= (\det \mathbf{F})^{-1/2}\exp \left[ -
\frac{1}{2}\xi \mathbf{F}^{-1}\xi \right] , 
\qquad \int W_{1}(\xi ) d^{2N} \xi/(2\pi)^{N} =1.
 \label{w3}
\end{equation}
Here $\mathbf{F}$ is a symmetric $2N\times 2N$ positively definite matrix.
The first-order mean values of the reservoir quadratures are supposed
to be equal to zero, for the sake of simplicity.
In particular, the state (\ref{w3})
may correspond to some mixed thermal state
or some pure squeezed vacuum state.
The dynamics of the subsystem $\mathbf{Q}$ is described by
the {\em reduced Wigner function\/}
$
W(\mathbf{Q},t)=\int W(\mathbf{Q},\xi,t )d^{2N}\xi /(2\pi )^{N}$,
which can be written as follows,
\be
 W(\mathbf{Q},t) = (2\pi  )^{-N} \int
d\mathbf{Q}^{\prime }d\xi^{\prime }d\xi \,
W_{0}(\mathbf{Q}^{\prime })W_{1}(\xi^{\prime })
\delta (\mathbf{Q}-\mathbf{R}_{11}
\mathbf{Q}^{\prime }-\mathbf{R}_{12}\xi^{\prime })
\delta (\xi -\mathbf{R}_{21}\mathbf{Q}^{\prime }-\mathbf{R}
_{22}\xi^{\prime }) .
\ee
The integration over $d\xi $ simply removes the second delta-function. To
perform the integration over $d\xi^{\prime }$ one should replace the first
delta-function with its integral representation 
$
\delta (\mathbf{x})=\int
e^{i\mathbf{kx}}d^{2N}\mathbf{k}/(2\pi )^{2N}$.
 Thus one arrives at two \emph{Gaussian\/} integrals (the first one over $d\xi^{\prime }$ and
the second over $d\mathbf{k}$), which can be easily
calculated exactly. The final result is the formula,
which has the same form as (\ref{w1}), but
with the variables $\mathbf{Q}$ and $\mathbf{Q}^{\prime }$ instead of
$\mathbf{q}$ and $\mathbf{q}^{\prime }$,
and with the
\emph{averaged propagator\/}
\be
G(\mathbf{Q},\mathbf{Q}^{\prime },t) = (2\pi )^{-n}[\det \mathcal{M}
_{\ast }(t)]^{-1/2}
\exp \left[ -\frac{1}{2}(\mathbf{Q}-\mathbf{R}_{11}
\mathbf{Q}^{\prime })\mathcal{M}_{\ast }^{-1}(\mathbf{Q}-\mathbf{R}_{11}
\mathbf{Q}^{\prime })\right] ,  
\label{w6}
\ee
where symmetric $2 \times 2$ matrix $\mathcal{M}_{\ast }(t)$ equals
\begin{equation}
\mathcal{M}_{\ast }(t)=\mathbf{R}_{12}(t)\mathbf{F}\widetilde{\mathbf{R}}_{12}(t).
\label{M*}
\end{equation}

A direct inspection shows that the propagator (\ref{w6}) (consequently, its
convolution with any initial function) satisfies the Fokker-Plank equation
\begin{equation}
\frac {\partial W}{\partial t}=-\frac {\partial}{
\partial Q_{\alpha}}\left[({\bf A}{\bf Q})_{\alpha}W\right
]+D_{\alpha\beta}\frac {\partial^2W}{\partial Q_{\alpha}\partial
Q_{\beta}}
\label{FPeq}
\end{equation}
with the following time-dependent drift matrix
$\mathbf{A}$
and time-dependent matrix of diffusion coefficients $\mathbf{D}$:
\begin{equation}
\mathbf{A}=\dot {\mathbf{R}}_{11}\mathbf{R}_{11}^{-1}
= \mathcal{A}_{11} + \mathcal{A}_{12}{\mathbf{R}}_{21}{\mathbf{R}}_{11}^{-1},
\label{A-R}
\ee
\be
2\mathbf{D}=\dot{\mathcal{M}}_{\ast }-\mathbf{A}
\mathcal{M}_{\ast }-\mathcal{M}_{\ast }\widetilde{\mathbf{A}}
= \mathcal{A}_{12}\left({\bf R}_{22} -{\bf R}_{21}{\bf R}_{11}^{-1}{\bf R}_{12}\right)
{\bf F} \tilde{{\bf R}}_{12}
+ {\bf R}_{12}{\bf F} \left(\tilde{{\bf R}}_{22} 
-\tilde{{\bf R}}_{21}\tilde{{\bf R}}_{11}^{-1}\tilde{{\bf R}}_{12}\right)
 \tilde{\mathcal{A}}_{12}.
\label{A-D}
\end{equation}

On the other hand, it is well known \cite{book,Lax,Haken} that the description of the evolution of an open quatum system
by means of the equation (\ref{FPeq}) is equivalent to the Heisenberg--Langevin equation
\be
d\hat {{\bf Q}}/dt=A\hat {{\bf Q}}+\hat{\chi }(t)
\label{176-10}
\ee
with delta-correlated in time noise operators
$\hat{\chi}_{\alpha}(t)$, 
\be\left\langle\hat\chi_{\alpha}(t)\right\rangle =0,
\qquad\left\langle\hat\chi_{\alpha}(t)\hat\chi_{\beta}(t')\right\rangle 
=\delta\left(t-t'\right)X_{\alpha\beta}.
\label{176-11}
\ee
The $2\times2$ noise matrix $\mathbf{X}$ and its transposition $\tilde{\mathbf{X}}$ are related to the diffusion matrix $\mathbf{D}$ as follows,
\be
4\mathbf{D}= \mathbf{X} + \tilde{\mathbf{X}}, \qquad
\mathbf{X} =
\left\Vert
\begin{array}{cc}
X_{11} & X_{12}
\\
X_{21} & X_{22}
\end{array}
\right\Vert.
\label{D-X}
\ee

In many applications (including the case considered here) the elements of the
interaction matrices ${\cal A}_{12}$ and ${\cal A}_{21}$ are {\em small}.
Then one can use the method of successive perturbations.  In the zeroth approximation 
the solution of Eq. (\ref{5.5}) reads
(as soon as matrix ${\cal A}_{22}$ is supposed to be time independent)
\begin{equation}
{\bf R}_{22}^{(0)}(t)=\exp\left({\cal A}_{22}t\right)
= \left\Vert\begin{array}{cc}
\mbox{diag}(\cos\omega_it)&\mbox{diag}(-\omega_i\sin\omega_it)\\
\mbox{~diag}(\omega_i^{-1}\sin\omega_it)&\mbox{diag}(\cos\omega_i
t)\end{array}
\right\Vert .
\label{5.9}
\end{equation}
Putting ${\bf R}_{22}^{(0)}(t)$ into the right-hand side of  (\ref{5.3}) one
obtains the first-order solution for the matrix ${\bf R}_{21}$:
\begin{equation}
{\bf R}_{21}^{(1)}(t)=\exp\left({\cal A}_{22}t\right)
\int_0^t\exp\left(-{\cal A}_{22}\tau\right){\cal A}_{21}(\tau){\bf R}_{11}^{(0)}(\tau)\,\mbox{d}\tau .
\label{5.10}
\end{equation}
Within the same accuracy, formula (\ref{A-D}) can be simplified as
\be
2\mathbf{D}= \mathcal{A}_{12}{\bf R}_{22}^{(0)} {\bf F} \tilde{{\bf R}}_{12}^{(1)}
+ {\bf R}_{12}^{(1)}{\bf F} \tilde{{\bf R}}_{22}^{(0)}  \tilde{\mathcal{A}}_{12}.
\label{A-D2}
\end{equation}

\section{Special case of very short interaction time}
\label{sec-short}

At this point it is necessary to take into account one of the most important results
of the previous theoretical studies on the DCE in cavities with semiconductor mirrors,
namely, that the recombination time of the carriers must be much shorter than the
period of the field oscillations of the fundamental cavity mode and the duration of each
laser pulse must be of the order of the recombination time or shorter. For the frequency
$\omega_0 \approx 2.5\,$GHz the period of oscillations is about $400\,$ps, while
the recombination time should not exceed $35\,$ps \cite{DD-rev2}.
This means that during the whole interval of time, when $\omega(t)$ is (slightly)
different from the initial (and final) value $\omega_0$ and the coupling coefficients are
different from zero, the matrix 
${\bf R}_{22}^{(0)}(t)=\exp\left({\cal A}_{22}t\right)$ 
in Eq. (\ref{5.10}), as well
as the matrix ${\bf R}_{11}^{(0)}(t)$ (although it cannot be written as a simple matrix exponential function
if the frequency $\omega$ depends on time), can be replaced by the {\em unity matrices}.
Of course, one has to make an additional (but quite realistic) assumption that the coupling coefficients with the
phonon modes with $\omega_i \gg \omega_0$ are negligibly small.

Then Eq.  (\ref{A-R})  leads to the following
first-order approximation for the drift matrix ${\bf A}$ governing the evolution of
the average values of the subsystem variables:  
\begin{equation}
{\bf A}^{(1)}(t)={\cal A}_{11}(t)+{\cal A}_{12}(t){\bf R}_{21}^{(1)}(t)
= {\cal A}_{11}(t)+{\cal A}_{12}(t)
\int_0^t{\cal A}_{21}(\tau)\,\mbox{d}\tau.
\label{5.11}
\end{equation}
Similarly, one obtains the approximate formula
$
{\bf R}_{12}^{(1)}(t) \approx \int_0^t{\cal A}_{12}(\tau)\,\mbox{d}\tau$,
so that the diffusion matrix  (\ref{A-D2}) can be written as
\be
2{\bf D} = {\cal A}_{12}(t){\bf F} \tilde {\bf R}_{12}^{(1)}(t)
+ {\bf R}_{12}^{(1)}(t) {\bf F} \tilde{\cal A}_{12}(t)
= {\cal A}_{12}(t) {\bf F} \int_0^t \tilde {\cal A}_{12}(\tau)\,\mbox{d}\tau
+ \int_0^t {\cal A}_{12}(\tau)\,\mbox{d}\tau  {\bf F} \tilde{\cal A}_{12}(t).
\label{5.17}
\ee

Taking into account the explicit form (\ref{5.8}) of matrices ${\cal A}_{12}(t)$ and ${\cal A}_{21}(t)$,
the elements of the dissipative part of the drift matrix 
\be
\mu(t) \equiv  {\bf A}^{(1)}(t) - {\cal A}_{11}(t)
= {\cal A}_{12}(t)\int_0^t{\cal A}_{21}(\tau)\,\mbox{d}\tau
\label{mu-A}
\ee
can be written as follows,
\beqnn
&&\mu_{11}(t) = \sum_{k=1}^N \int_0^t d\tau\left[v_k(t)u_k(\tau) - g_k(t)z_k(\tau)\right],
\qquad
\mu_{12}(t) = \sum_{k=1}^N \int_0^t d\tau\left[v_k(t)g_k(\tau) - g_k(t)v_k(\tau)\right],
\\
&&\mu_{21}(t) = \sum_{k=1}^N \int_0^t d\tau\left[u_k(t)z_k(\tau) - z_k(t)u_k(\tau)\right],
\qquad
\mu_{22}(t) = \sum_{k=1}^N \int_0^t d\tau\left[u_k(t)v_k(\tau) - z_k(t)g_k(\tau)\right].
\eeqnn

In the case of the DCE with a semiconductor mirror, the time dependence of the coupling coefficients 
between the field and the ``reservoir'' is introduced in order to take into account the
time dependence of the number of carriers inside the semiconductor (since there is no coupling
in the absense of carriers, if the quality factor of the empty cavity is big enough).
Therefore I assume that {\em all\/} coupling coefficients
are proportional to {\em the single\/} time dependent factor $\nu(t)$ 
(related somehow to the number of carriers):
\be
u_k(t)= \nu(t)U_k, \quad
v_k(t)= \nu(t)V_k, \quad
g_k(t)= \nu(t)G_k, \quad
z_k(t)= \nu(t)Z_k,
\label{uU}
\ee
where the coefficients $U_k, V_k, G_k, Z_k$ do not depend on time.
Then straightforward simple calculations lead to the following remarkable result:
\be
\mu_{12} = \mu_{21} \equiv 0, \qquad
\mu_{11} \equiv \mu_{22} = \lambda(t)\sum_{k=1}^N \left(U_k V_k - G_k Z_k\right),
\qquad \lambda(t) = \nu(t)\int_0^t \nu(\tau) d\tau.
\label{mu-good}
\ee
Consequently, two equal damping coefficients in Eq. (\ref{Fx-Fp}) arise automatically:
$\gamma_x \equiv \mu_{22} =\mu_{11} \equiv \gamma_p$.

If the reservoir variance matrix has the diagonal form
${\bf F}=\left\Vert\begin{array}{cc}
\mbox{diag}\left(\omega_i^2f_i\right)&0\\
0&\mbox{diag}\left(f_i\right)\end{array}
\right\Vert$ 
(in particular, ${\bf F}$ may be an equilibrium variance matrix for the 
reservoir variables), then Eqs. (\ref{5.8}), (\ref{5.17}) and (\ref{uU})
result in the following expressions for the elements of the $2\times2$ symmetric diffusion matrix ${\bf D}$:
\be
D_{11}(t) = \lambda(t)\sum_{k=1}^N f_k\left(\omega_k^2 V_k^2 + G_k^2\right),
\qquad
D_{22}(t) = \lambda(t)\sum_{k=1}^N f_k\left(\omega_k^2 Z_k^2 + U_k^2\right),
\label{D22}
\ee
\be
D_{12}(t) = -\lambda(t)\sum_{k=1}^N f_k\left(\omega_k^2 V_k Z_k + G_k U_k\right).
\label{D12}
\ee
These coefficients become especially simple if the interaction Hamiltonian in (\ref{5.1}) has the 
RWA form:
\be
\hat{H}_{int} = \nu(t)\sum_{k=1}^N \left(\rho_k \hat{a}_0\hat{a}_k^{\dagger} + \rho_k^* \hat{a}_k\hat{a}_0^{\dagger}\right),
\qquad \hat{a}_0=\frac{\omega_0 \hat{x}_0 +i\hat{p}_0 }{\sqrt{2\omega_0}},
\quad \hat{a}_k=\frac{\omega_k \hat{x}_k +i\hat{p}_k }{\sqrt{2\omega_k}},
\label{RWA}
\ee
where $\omega_0$ is the initial frequency of the central oscillator.
Then
\be
G_k= \sqrt{\omega_0\omega_k}\,\mbox{Re}(\rho_k),
\quad Z_k= \frac{\mbox{Re}(\rho_k)}{\sqrt{\omega_0\omega_k}},
\quad V_k = \sqrt{\frac{\omega_0}{\omega_k}}\,\mbox{Im}(\rho_k),
\quad U_k = -\sqrt{\frac{\omega_k}{\omega_0}}\,\mbox{Im}(\rho_k),
\label{g-u-RWA}
\ee
so that
$D_{12} \equiv 0$ and $D_{11} =\omega_0^2 D_{22} = \lambda(t)\omega_0
\sum_{k=1}^N f_k \omega_k |\rho_k|^2$.
Using (\ref{D-X}) one arrives at the set of the noise coefficients which has
exactly the same structure as the {\em minimum noise\/} set of coefficients
introduced in \cite{DD-rev2,D-job05} in the framework of a phenomenological approach:
\be
\gamma_x=\gamma_p=\gamma(t), \qquad
\chi_{xp} = - \chi_{px}= i\gamma(t), \qquad
\chi_{pp}= \omega_0^2\chi_{xx}=\gamma(t) \omega_0 G.
\label{min-sym}
\ee
The same set (\ref{min-sym}) with $\gamma=const$ arises in the case
of the RWA coupling with {\em time-independent\/} coefficients for $\omega_0 t \gg 1$ \cite{DMM95}. 
This observation makes the ``bridge'' between the short-time and long-time limits.
It supports the assumption made in \cite{DD-rev2,D-job05} that the coefficient $G$ can be identified
with the mean photon number in the equilibrium state with the given temperature $T$ and frequency $\omega_0$,
$G=\coth\left[\hbar\omega_0/(2k_B T)\right]$.

\section{Conclusion}

It was shown that the ``symmetric minimum noise''  
set of coefficients in the Heisenberg--Langevin equations for 
a quantum damped nonstationary oscillator, used in the studies of the dynamical
Casimir effect in a cavity with a photo-excited semiconductor mirror,
arises quite naturally from any bilinear interaction between the selected field
mode and the {\em bosonic\/} bath. The main conditions are: (i) the duration of interaction is 
much shorter than the period of the oscillator and (ii) all the coupling
coefficients are proportional to a single function of time.
Of course, real interactions are much more complicated, so the construction of a more realistic model
remains to be a very interesting and important problem. Nonetheless, the results obtained
give a strong support to the conclusions of recent studies \cite{DD-rev2,DCas60,D-PRA09} 
concerning the feasibility of the MIR 
experiment and the necessary conditions for its success.

\section*{Acknowledgements}
\pst
A partial support from CNPq (DF, Brazil) is acknowledged.

\end{document}